\documentclass[preprint]{aastex}

\shorttitle{Fast Integrated Spectra Analyzer}

\shortauthors{Ben\'itez-Llambay et al.}

\begin{document}

\title{Fast Integrated Spectra Analyzer: A New 
Computational Tool For Age and Reddening Determination of Small Angular 
Diameter Open Clusters}

\author{ALEJANDRO BEN\'ITEZ-LLAMBAY}
\affil{Instituto de Astronom\'ia Te\'orica y Experimental, Observatorio 
Astron\'omico de C\'ordoba, Universidad Nacional de C\'ordoba,
Laprida 854, 5000, C\'ordoba, Argentina}
\email
{alejandrobll@oac.uncor.edu}

\author{JUAN J. CLARI\'A}
\affil{Observatorio Astron\'omico, Universidad Nacional 
de C\'ordoba,
Laprida 854, 5000, C\'ordoba, Argentina}
\email{claria@oac.uncor.edu}

\and

\author{ANDR\'ES E.
 PIATTI}
\affil{Instituto de Astronom\'{\i}a y F\'{\i}sica del Espacio, CC 67, Suc.
28, 1428, Ciudad de Buenos
 Aires, Argentina}
\email{andres@iafe.uba.ar}

\begin{abstract}

We present a new algorithm called 'Fast Integrated Spectra Analyzer" (FISA) that permits fast and 
reasonably accurate age and reddening determinations for small angular diameter open clusters by using 
their integrated spectra in the (3600-7400) \AA \ range and currently available template spectrum 
libraries. This algorithm and its implementation help to achieve astrophysical results in shorter times 
than from other methods. A brief review is given of the integrated spectroscopic  technique applied to the 
study of open clusters as well as the basic assumptions that justify its use. We describe the 
numerical algorithm employed in detail, show examples of its application, and provide a link to the code. Our method 
has successfully been applied to integrated spectroscopy of open clusters, both in the Galaxy and in the Magellanic Clouds, 
to determine ages and reddenings.
\end{abstract}


\section{Introduction}

One of the main questions posed by stellar astrophysics has been, and continues to be, the formation and 
evolution of the Milky Way Galaxy. Thanks mainly to many determinations of chemical abundances, both for 
individual stars and for stellar systems in the Galaxy, remarkable advances in the understanding of these 
processes have taken place during the last four decades. However, in spite of these advances, a variety 
of issues related to Galactic evolution are not yet fully understood. These include such topics as the history 
of stellar formation, the initial mass function, and the processes of chemical enrichment. Any way to increase 
our knowledge of these subjects through observations of astronomical objects is obviously justified.

Observations of star clusters can be particularly useful. First, it is common knowledge that the Galactic 
open and globular clusters cover most of the age and chemical abundance ranges known in the universe, and one 
can therefore consider the Galactic cluster system as representative of the general stellar population in terms 
of age and metallicity. Second, the determinations of the fundamental parameters of open clusters (OCs) has undeniable 
importance in better understanding the structure and chemical evolution of the Galactic disk.

In general, such parameters as interstellar reddening, distance, age and even metallicity may be determined more 
easily and more accurately for star clusters than for isolated stars. The determination of  
such parameters in OCs has usually been based  on color-magnitude diagrams (e.g., Lyng\aa\, 1987; Hasegawa et 
al. 2008) and/or on photometric and spectroscopic studies of individual stars (e.g., Clari\'a et al. 2006; Mermilliod 
et al. 2001). However, for small angular diameter OCs, the integrated spectra can be used to infer those 
properties (e.g., Santos \& Bica 1993). Observations of cluster integrated spectra have been carried out at the 
Observatorio Astron\'omico de C\'ordoba (Argentina) over the past decade. Analysis of some of these data has 
demonstrated that this is a good way to determine cluster parameters with comparatively short observing times (see, e.g., 
Ahumada et al., 2000, 2001, 2002, 2007; Piatti et al. 2005; Palma et al. 2008a,b; Talavera et al. 2010).

Given that the integrated spectra technique represents an efficient way to increase the amount of observational 
data for OCs, we want to have a method that systematically and automatically analizes  such data and yields 
astrophysical results. This article describes the development of the algorithm used to develop our computational 
tools to analyze the sample of previously unstudied small angular diameter Galactic OCs, which have been 
observed at the Complejo Astron\'omico El Leoncito (CASLEO, Argentina) to determine ages and reddenings.

This paper is organized as follows: Section 2 discusses the main ideas that justify the use of integrated 
spectroscopy to estimate ages and reddenings of small angular diameter OCs. Section 3 first describes 
two different methods of determining the $E(B-V)$ color excess that affects a star cluster's integrated 
emitted flux. We next show how these two methods can be combined in order to simultaneously determine the age and 
reddening of an OC. Finally, we present in this section a new algorithm,''Fast Integrated Spectra Analyzer" (FISA), 
that is suitable for a fast determination of the two mentioned OC parameters. Section 4 gives two examples of 
how FISA works in different cases. Section 5 summarizes our main conclusions.

\section{Stellar population ages}

It is currently possible to compute in detail the spectral energy distribution of a star from the 
theory of stellar atmospheres. If we consider a certain number of stars in a given area of space and we 
know the mass, the chemical composition and the evolutionary stage of each star, it is possible to determine 
the spectrum of the light emitted by each star and, therefore, what the integrated spectrum of all the stars 
will be. This approach to determining the energy output for astronomical sources in which it is impossible to 
resolve their individual components is called stellar population synthesis theory (see, e.g., Cid Fernandes et al. 
2005 and references therein). Typically this theory is used in interpreting the spectra of galaxies whose stars are 
not resolved.

For a stellar cluster one must take into account that the distribution of stellar energies changes through time. 
The more massive stars leave the main sequence before less massive ones and, in the course of time, this causes a decrease 
in the number of blue stars in the cluster relative to the red ones. Thus, the integrated spectral energy distribution 
of a cluster constitutes a parameter that changes over time, gradually becoming redder, and it seems reasonable that 
this parameter can be an indicator of cluster age.

One can hypothesize that at any point in its history a cluster is characterized solely by its integrated light. 
This leads to the idea of building a library of integrated spectra templates for star clusters in different 
evolutionary stages that can be used for interpreting observations. There are two ways of doing so. The first consists 
of building the templates synthetically based on the stellar evolution theory (e.g., Bessell et al. 1998, Kurucz 1992). 
The second way consists of observing a large number of OCs whose ages were determined by direct methods (e.g., using 
color-magnitude diagrams) and, from these, developing the library of template spectra (e.g., Santos et al. 1995, Piatti et 
al. 2002; Ahumada et al. 2007).

\subsection{Stellar Population Synthesis}

The physical properties of a star at any given stage in its life can be described by an appropriate stellar model that can be 
computed if the star's initial mass and metallicity are known. As we can only measure the radiative fields coming 
from the stars, luminosity and effective temperature seem to be the most appropriate theoretical properties on which 
to focus. A star's evolution is represented by an evolutionary track in the $L$ versus $T_{e}$ diagram. These two parameters 
are related to absolute magnitude and color, respectively, and therefore evolutionary tracks provide the basis for 
understanding the observed Hertzsprung-Russell (HR) diagrams.

There are many computational codes available in the astronomical literature for computing stellar models and stellar 
evolution under different assumptions and deriving evolutionary tracks of stars in the $L$ versus $T_{e}$ diagram. The 
tracks are often combined to obtain isochrones that represent the locus in the diagram for a group of stars that 
formed together. Examples can be seen in \citet{ls01} and \citet{getal02}.

According to stellar population synthesis theory, the energy distribution of an unresolved star cluster will be the sum of the 
individual contributions of each star. There are essentially two different approaches to determining a cluster's age by using 
integrated spectroscopy. One is to compare the integrated radiation with the predictions of stellar evolution theory. 
These so-called evolutionary population synthesis methods aim to reproduce the observed spectrum by combining information 
for the entire stellar system using libraries of evolutionary tracks and stellar spectra with prescriptions for the initial mass 
function (IMF), star formation rate, and chemical histories. The main advantage of this approach is the possibility of 
progressing through a large group of parameters to generate synthetic spectra for all the required conditions. In practice, 
very sophisticated stellar atmospheres codes are used (e.g., Bessell et al. 1998; Kurucz 1992). Once a synthetic spectrum library 
is available, the integrated spectrum of a studied cluster can be fit by making a linear combination of the library's synthetic 
spectra. 

The second approach is to apply empirical population synthesis methods. These aim at reproducing the observations by means of 
linear combinations of observed spectra of individual stars, of individual clusters, etc. Three different empirical methods 
have been used to try to characterize cluster age by using integrated spectroscopy:

\begin{enumerate}
\item{The first method consists of obtaining spectra of nearby stars with known bolometric magnitudes $M$$_b$, 
effective temperatures $T$$_e$, and metallicities $Z$. Once these spectra are available, the particular spectrum of 
a star having other $M$$_b$, $T$$_e$ and $Z$ values can be determined by means of a simple interpolation. Spectral libraries 
suitable for this can be seen in, for example, \citet{gs83}, \citet{s98} and \citet{lbetal03}. This technique can also be 
used to characterize a stellar population - using linear combinations of the interpolated spectra - and it has been used 
to determine age and metallicity of OCs (e.g., Dias et al. 2010). The method is somewhat limited, however, 
because good-quality spectra with well-determined parameters are available only for stars in the solar 
neighborhood, which constitute a very limited stellar sample.}
\item{The second empirical method for characterizing a cluster age by using integrated spectroscopy was explored by 
\citet{ba86}. They found that the equivalent widths of the Balmer lines in a cluster integrated spectrum are related 
to its age. They showed, however, that in most cases there is not a unique relationship between these two parameters. 
For this reason this equivalent width method is not reliable to fully characterize a cluster age. Nevertheless, 
this method is of interest because equivalent width determinations are practically reddening independent, which permits 
a cluster age to be estimated without knowing the reddening. }
\item{The third approach consists of spectroscopically observing OCs to obtain a library of template spectra from 
clusters of known ages. Because a cluster's integrated energy distribution is determined by its evolutionary stage, 
the ages of clusters can be found by comparing the continuum shape and the strength of the various 
absorption lines in their integrated spectra with these features in the integrated spectra of OCs of known ages. It was 
this third approach that inspired \citet{sb93} to start building a library of template spectra suitable for characterizing  
the evolutionary stages of Galactic and extragalactic clusters. It should be noted that, assuming the  Galactic 
IMF is universal, it is not necessary to make any assumption about its form, because the IMF is essentially built into 
the template spectra.}
\end{enumerate}

Based on stellar population synthesis theory and on the arguments previously stated, we may conclude that 
there must be spectral properties inherent to different stellar populations. These properties are usually called 
''integrated spectral properties,'' since they reflect the properties inherent to a given population group as a whole.

During the last 10 years, many studies have been published that present results from using the preceding third method to 
investigate a good number of Galactic OCs (Ahumada et al. 2000, 2001, 2007; Palma et al. 2008a) and Magellanic Cloud star 
clusters (Ahumada et al. 2002; Piatti et al. 2005; Santos et al. 2006; Clari\'a et al. 2007; Palma et al. 2008b; Talavera 
et al. 2010).

In the present article, the aforementioned third method for determining cluster ages will be adopted to derive cluster 
ages. 

\subsection{Library of Template Spectra}

Our approach to estimating ages of OCs from their integrated spectra is largely based on the third empirical 
method described in \S 2.1. We therefore present the basis on which we develop our method of age determination 
using template spectrum libraries of OCs.

Star cluster integrated spectra represent a very important link between the study of stellar populations of galaxies and 
individual stars. Seen in this way, observations of the integrated radiation of unresolved stellar sources have an 
undeniable significance. In addition, the determination of reddenings, ages and even metallicity of clusters can be 
performed more easily and more accurately than for individual stars.
 
\citet{pietal02} constructed a library of OC integrated spectra representing stellar populations of different ages. The 
method they used consisted of combining reddening-corrected spectra of clusters of practically the same age $\tau$. i.e., 
those within a small interval of age $\Delta \tau$. The individual spectra were averaged, weighting them according 
to the square of their signal-to-noise  ratios (S/N), to produce a single template spectrum having a good S/N. 
When each template was built, it was  ensured that the continuum distribution and the spectral absorption features of the 
cluster spectra  used to create such a template were very similar. In this way, a set  
of template spectra characterized by their different ages $\tau \pm \Delta \tau$ was obtained.

The result of \citet{pietal02} was 20 solar-metallicity template spectra in a wide age range, from those 
associated with gas emission until ages as old as a few billon years, generated from 47 integrated spectra of Galactic OCs  
observed mostly at CASLEO (Argentina). All the templates cover the same spectral range between 3600 \AA \ and 7400 \AA. 
Two new solar-metallicity templates corresponding to age groups of (4-5) Myr and 30 Myr were later created by 
Ahumada et al. (2007), while the templates for 20 Myr and for 3.5 Gyr were redefined using only spectra of high S/N, 
improving 
their quality. With these additions and improvements one now has available 22 high-S/N templates representing 
cluster evolutionary stages of 2-4, 4-5, 5-10, 20, 30, 40, 45-75, 100-150, 200-350, 500, 1x10$^{3}$ and 3-4x10$^{3}$ 
Myr. Table 1 lists the notation for the available template spectra along with the number of 
members associated with it. The first letter of the template indicates whether it corresponds to a young-age group (Y) or to an 
intermediate-age group (I). Within both the young- and intermediate-age groups, Piatti et al. (2002) used a second alphabetic 
character to identify different spectra and also included a third numeric character in the 2-4 and 5-10 Myr templates
 to differentiate cluster spectra corrected by different reddening values.

Following the procedure described by Clari\'a (2008), reddenings and ages of more than 50 small angular diameter 
Galactic OCs have been estimated (Ahumada et al. 2000, 2001, 2007, 2010; Palma et al. 2008a). Although 
metallicity has an impact on the spectra of OCs older than $\sim$ 0.5 Gyr, the metallicity features in the spectra of 
these clusters in the 3600 \AA \ $< 
\lambda <$ 7400 \AA\  spectral range are relatively weak (see, e.g., Bica \& Alloin 
1986), so they do not significantly affect the determination of the two derived parameters. The results are mainly 
based on comparison of the observed spectra with template spectra, but there was still some use of the 
relation between equivalent width and age found by \citet{ba86}. This was because the task of comparing an 
integrated spectrum with each of the templates in a library often turned out to be very time-consuming and difficult 
when there was no previous hint on the cluster's age. As will be seen in \S 3, the method of \citet{c08} is based 
on the minimization of a certain function. Measurement of the equivalent widths containing the Balmer spectral features 
suggested where to start looking for this minimum, thus constraining the search. The age finally adopted for a studied 
cluster was, however, only the result from the template match method. In the procedure described in this article, we managed 
to avoid this step by automating the search for the template that best matches the reddening-corrected observed spectrum.

To summarize, in this section we have tried to demonstrate why it is reasonable that the spectrum radiation distribution 
of a particular stellar population can reflect its evolutionary stage. OCs may be considered pure stellar populations 
in the sense that all their stars were formed at the same time and have just about the same metallicity. Thus, it seems 
reasonable to suppose that their integrated spectra may provide information about their ages. In fact, when one deals 
with unresolved sources it is only through their integrated spectra that one can investigate the astrophysical nature 
of the objects. Several examples of astrophysical applications of the integrated spectroscopic technique have been 
described by \citet{c08}. Thus far we have described three different ways to determine the ages of stellar populations by 
using only their integrated light. We will now describe how to estimate both ages and reddenings for OCs from integrated 
spectroscopy.

\section{Age and Reddening Determination}

It has been shown that it is possible to characterize the evolutionary stage of an OC from its integrated spectrum. 
The spectrum, of course, may be affected by interstellar reddening and we will now show that it is possible to 
derive OCs' reddenings using their integrated spectra. However, to determine both a cluster's age and its reddening 
through comparison with a template spectrum library implies finding a single fitting parameter, which we will call 
the "age-reddening" parameter. Physically, a cluster's age is independent of its reddening, but when a template spectrum 
library is used, one must find the most probable combination of both parameters (age and reddening). If the apparent 
interdependence between these two parameters is taken into account, one can establish a sound criterion for 
determining both age and reddening for OCs by using integrated spectroscopy. 

\subsection{Reddening Determination}

In this section, we describe a way of determining the $E(B-V)$ color excess affecting the integrated emitted flux 
of a star cluster by a procedure we call the ''average method.'' This method allows us to automate 
age and reddening determinations for OCs.

\subsubsection{The Average Method}

The average method consists of being able to find a mean cluster reddening in the sense that will be 
explained later. This method demands fewer operations than other procedures and yields good results. The 
fundamental ideas are described next.

Let us consider a light source (star or star cluster) located somewhere in the Galaxy. The observed monochromatic 
flux at a given wavelength $\lambda$  will be

\begin{equation}
 F(\lambda)=F_0(\lambda)e^{-\gamma_{\lambda}E(B-V)},
\end{equation}

\noindent wherein $F_0(\lambda)$ is the object's original flux, $\gamma_{\lambda}=-0.4 \ln{(10)} \xi(\lambda)$, 
and $E(B-V)$ is the color excess. In this work $\xi(\lambda)$ is the \citet{s79} interstellar extinction law.

For practical reasons, the flux is usually normalized by dividing it by the flux in a certain 
wavelength. Thus, the normalized equation renders

\begin{equation}
 F'(\lambda)=F'_0(\lambda)e^{-\gamma'_{\lambda}E(B-V)},
\end{equation}

\noindent Where it is now $F'(\lambda)=F(\lambda)/F(\lambda_0)$, $F'_0(\lambda)=F_0(\lambda)/F'_0(\lambda_0)$ and 
$\gamma'_{\lambda}=\gamma_{\lambda}-\gamma_{\lambda_0}$

From the preceding equation, it can be noted that if the object's original flux is known, it is possible to find $E(B-V)$ 
and viceversa. Assuming we know $F'_0(\lambda)$ then

\begin{equation}
 E(B-V)=-\displaystyle\frac{1}{\gamma'_{\lambda}} \ln G_{\lambda},
\end{equation}

\noindent wherein $G_{\lambda}=F'(\lambda)/F'(\lambda_0)$

The left side of equation (3) can be taken to be a constant. The first factor on the right side certainly is not; consequently, the variations of $G_{\lambda}$ and $\gamma'_{\lambda}$ should be such that the $\ln G_{\lambda}$ to 
$\gamma'_{\lambda}$ ratio is constant: i.e., equals - $E(B-V)$.

The noise in the flux measurements produced by photon counting affect the value of $E(B-V)$ 
determined in this way. It is therefore appropriate, from a physical point of view, to write 
 equation (3) in the following way:

\begin{equation}
 E(B-V)_i=-\displaystyle\frac{1}{\gamma'_{\lambda_i}} \ln G_{\lambda_i}.
\end{equation}

This equation shows that a different $E(B-V)_i$ value corresponds to each $\lambda_i$. Moreover, as the 
difference between the reddenings associated with various wavelengths are only the result of  noise introduced 
during the measuring process, it is expected that such reddenings are random and there will be a predictable reddening 
distribution. We now establish what this distribution is like.

\subsubsection{Color Excess Distribution}

It is well known that, for each $\lambda$, the noise associated with the observed flux follows a Poisson 
distribution since the flux measurement is made by counting photons. On the other hand, for very long integration 
times, the expected frequency of photon detection is very large. So the noise should resemble a normal 
distribution. Therefore, it is reasonable to accept that the observed flux belongs to a Gaussian distribution 
with a mean $\mu_{\lambda}$ and a standard deviation $\sigma_{\lambda}$. This information should be enough to 
fully characterize the color excess distribution. However, we will adopt a more intuitive way to derive it.

Assuming a theoretical reddened flux (without noise), the theoretical reddening  from equation (3) must be:

\begin{equation}
 E(B-V)_0=-\displaystyle\frac{1}{\gamma'_{\lambda}} \ln G_{\lambda},
\end{equation}

As mentioned, equation (5) is only true theoretically. However, if an actual $G_{\lambda}$ determination has 
an associated uncertainty $\delta G_{\lambda}$, then equation (5) can be written as

\begin{equation}
 E(B-V)_{\lambda}=-\displaystyle\frac{1}{\gamma'_{\lambda}} \ln \left [ G_{\lambda} + \delta G_{\lambda} \right ].
\end{equation}

Assuming the observations have a reasonably good S/N, we expect $\delta G_{\lambda} \sim 0$. It is then possible 
to express equation (6) to first order as 

\begin{equation}
 E(B-V)_{\lambda}=E(B-V)_0-\displaystyle\frac{\delta G_{\lambda}}{G_{\lambda}}\displaystyle\frac{1}{\gamma'_{\lambda}}.
\end{equation}

So, in this approach, the variations in measured $E(B-V)$ values come only from the measurement process and follow the 
adopted extinction curve.

In the visual spectral range, the $\gamma_{\lambda}$ extinction curve is proportional to $1/\lambda$ so that the 
most probable reddening for each $\lambda$ is

\begin{equation}
 E(B-V)_{\lambda}=E(B-V)_0-\displaystyle\frac{\delta G_{\lambda}}{G_{\lambda}}\displaystyle\frac{\lambda \lambda_0}
 {\alpha(\lambda_0-\lambda)}, 
\end{equation}

\noindent where $\alpha$ is the proportionality constant whose value is $\sim 10^{4}$ \AA \ mag$^{-1}$. Thus, we have 
shown that, for each wavelength, the $E(B-V)$ value we expect to measure rises in proportion to 
$(\delta G_{\lambda} \lambda_0)/(\alpha G_{\lambda})$ and follows the $\lambda/(\lambda_0-\lambda)$ curve.

A good way of applying this method is by defining a mean reddening $<E(B-V)>$. This procedure is justified 
as follows. The mean reddening is defined as

\begin{equation}
 <E(B-V)>=\lim_{t \to \infty}{\displaystyle\frac{1}{2t} \displaystyle\int_{-t}^{t} E(B-V)_\lambda d\lambda}.
\end{equation}

Although it might be expected that the $\delta G_{\lambda} / G_{\lambda}$ ratio slightly depends on the wavelength 
due to variations in instrumental sensitivity, to simplify the discussion we assume the ratio is independent of 
$\lambda$. It is then evident that, in the absence of systematic errors, $\delta G_{\lambda}$ will cause the measured 
points to be above or below the $1/\gamma'_{\lambda}$ curve by chance. The associated $1/\gamma'_{\lambda}$ curve depends 
as much on the sign of $\delta G_{\lambda}$ as on its absolute value. Owing to this fact, two 
consecutively measured points may be situated on fairly different curves if the sign of $\delta G_{\lambda}$ changes. 
The consequence is that the measured points will populate a dense region in the $E(B-V)_{\lambda}$ versus $\lambda$ diagram, 
corresponding to all possible curves proportional to $1/\gamma_{\lambda}$, with their proportionality constants
 depending on the normalizing wavelength, on the $\alpha$ value and on the $\delta G_{\lambda}/G_{\lambda}$ ratio. 
This is illustrated in Figure 1, where we compute $E(B-V)$ values using a spectrum that has been reddened and had noise added 
as shown in Figure 2. It is clear that the measured points do populate the region of the allowed $1/\gamma'_{\lambda}$ 
curves. The average reddening over all points is 0.5, which agrees with the reddening that was applied to the original 
spectrum.

Since $\delta G_{\lambda}$ is a random variable and the absolute value of $\delta G_{\lambda}/G_{\lambda}$ 
should suffer little change in a neighboring continuum, it is expected that the points measured on a determined 
curve proportional to $1/\gamma'_{\lambda}$ cancel with their opposite. In this way, they produce the net effect that 
the mean value of all points is $E(B-V)_0$. This occurs when we have the same number of points on both sides 
of the normalizing point. So, in absence of systematic errors, we expect that

\begin{equation}
 \lim_{t \to \infty} \displaystyle\frac{1}{2t}\displaystyle\int_{-t}^{t}\displaystyle\frac{\delta 
 G_{\lambda}}{G_{\lambda}} \displaystyle\frac{\lambda}{\lambda_0- \lambda} d\lambda = 0.
\end{equation}

In the case of a discrete distribution, the mean reddening  is

\begin{equation}
 <E(B-V)>=\displaystyle\frac{1}{N} \displaystyle\sum_{i=1}^{N} { E(B-V)_i },
\end{equation}

\noindent where the subindex $i$ indicates the measured reddening using the $\lambda_i$ wavelength according 
to equation (4) and $N$ is the number of analyzed spectral points.

The basic algorithm to determine reddening according to this method is the following: (i) Normalize both the 
original and the observed spectra at the mean value  $<\lambda>=(\lambda_{max}-\lambda_{min})/2$  to ensure that the 
number of points lying on both sides of the normalizing point are the same. (ii) Compute the extinction curve 
$\gamma'_{\lambda}=\gamma_{\lambda}-\gamma_{<\lambda>}$. (iii) Determine the reddening for each $\lambda$ according to equation (4). (iv) Determine $<E(B-V)>$ according to equation (11) and adopt the resulting value as the 
reddening affecting the spectrum.

In the next section, we describe another method to determine the reddening affecting an astronomical object. 
This one, together with the average method, will provide a powerful tool to simultaneously determine age and reddening of 
small angular diameter star clusters using integrated spectroscopy.

\subsubsection{The $\chi^2$ Method}

Let us again consider equation (2) which relates the observed flux of an object with its reddening, its original 
flux and the extinction law.

The original or intrinsic flux of the object is then

\begin{equation}
 F'_0(\lambda)=F'(\lambda)e^{\gamma'_{\lambda}E(B-V)}.
\end{equation}

Since the previous equation bears the same form as equation (2), it might be thought that to find $E(B-V)$, we 
can compute a sequence of fluxes whose terms result from the application of different reddenings to the observed 
flux. The basic idea consists of calculating as many cases as possible until one observed flux turns out to be 
equal to the object's original flux, assuming that the latter is known or given by a template spectrum. We can 
then construct the following sequence:
\begin{eqnarray}
 F'(\lambda)&=&F'_0(\lambda)e^{-\gamma'_{\lambda}E(B-V)} \\ \nonumber
 F'_1(\lambda)&=&F'(\lambda)e^{-\gamma'_{\lambda}E(B-V)_1} \\ \nonumber
 F'_2(\lambda)&=&F'(\lambda)e^{-\gamma'_{\lambda}E(B-V)_2} \\ 
 &\vdots& \nonumber \\ \nonumber
 F'_m(\lambda)&=&F'(\lambda)e^{-\gamma'_{\lambda}E(B-V)_m}. \\ \nonumber 
\end{eqnarray}

If we find any $E(B-V)_j$ value for which we obtain $F_j(\lambda)=F_0(\lambda)$, we will have solved the problem 
of finding the $E(B-V)$ color excess affecting the observed flux. Indeed, $E(B-V) = -E(B-V)_j$.

The most direct way to determine if $F_j(\lambda)$ is equal to the unreddened object original flux is to analyze 
the following normalized quadratic sum of the residuals:

\begin{equation}
 \chi^2(E(B-V)_j)=\displaystyle\sum_{i=1}^N \displaystyle\frac{(F'_j(\lambda_i)-F'_{0}(\lambda_i))^2}{F'_{0}(\lambda_i)}.
\end{equation}

Function $\chi^2$ should have a minimum value within a closed reddening interval. We will not show here 
that this minimum is global and unique. Rather, we must justify in some way in which $E(B-V)$ domain it should 
be sought. This will be explained later, when we give details about the method finally adopted for reddening 
determination. For now, the algorithm to find $E(B-V)$ following the $\chi^2$ method is this: (i) Normalize both the observed 
and the original spectra in a given $\lambda_0$. (ii) Compute the extinction curve 
$\gamma'_{\lambda}=\gamma_{\lambda}-\gamma_{\lambda_0}$. (iii) Build up the flux sequence equation (13) by de-reddening 
the observed flux by different $E(B-V)_j$ values. (iv) Compute the $\chi^2(E(B-V)_j)$ function. (v) Adopt the $-E(B-V)_j$ corresponding to the minimization of the $\chi^2(E(B-V)_j)$ function as most probable 
reddening value.

\subsection{Simultaneous Age and Reddening Determination}

We have shown in previous sections how the reddening that affects an astronomical object can be determined 
using integrated spectroscopy. We assumed, however, that the object's original flux was already known. In this section, we 
will see that a reddening determination cannot be performed independently from the determination of the original flux. From a 
mathematical point of view, reddening depends on the original flux and viceversa, while from the physical point of 
view there is no such degeneracy. A criterion must be established that allows an unambiguous determination of both the object's 
intrinsic flux and the $E(B-V)$ color excess affecting it.

When OCs are considered, as shown in \S 2, the determination of the original flux can be carried out by 
using template spectra, which in turn leads to estimates of the cluster's age and reddening. But this is true only if the 
adopted extinction law is the correct one for the cluster's Galactic direction. We will next show how incorporating 
the average and $\chi^2$ methods of reddening determination allows one to simultaneously  determine the age and reddening 
of an OC.

\subsubsection{Combining the Average and $\chi^2$ Methods}

If we believe that the observed flux of an object is related to an $E(B-V)$ that is well defined and unique for the 
object based on an erroneously selected $k$th template spectrum from:

\begin{equation}
 F'(\lambda)=F'_{0k}(\lambda) e^{-\gamma'{\lambda} E(B-V)},
\end{equation}

\noindent then there will be a systematic error in the  individual $E(B-V)_{\lambda}$ determinations. Indeed, in 
this case and in the absence of noise, equation (3) will not be valid, because a relation will not exist between 
$G_{\lambda}$ and $\gamma'_{\lambda}$ that permits $E(B-V)$ to be a constant. So, even if the noise in the observed 
flux is small, $E(B-V)_{\lambda}$ will not follow the distribution given by equation (8), because the 
large systematic component in $G_{\lambda}$ will distort it. The  distribution will become increasingly different from the 
one theoretically found as the difference between the $k$th template spectrum and the object's original flux 
increases. This behavior, illustrated in Figure 3, provides a way to ascertain if the adopted template spectrum is 
correct. If the observed reddening distribution can be matched by the one theoretically found, within certain limits, 
we may very well think that the adopted template is the right one. In case this is impossible to achieve, the 
template must be discarded and, consequently, the $E(B-V)$ value determined through the ''average method'' will not 
have any physical sense.

When the correct template spectrum is unknown, the $\chi^2$ method demands finding the minimum value of the 
following function:

\begin{equation}
\chi^2(E(B-V)_j;F'_{0k})=\chi_{jk}^2=\displaystyle\sum_{i=1}^N 
\displaystyle\frac{(F'_j(\lambda_i)-F'_{0k}(\lambda_i))^2}{F'_{0k}(\lambda_i)}, 
\end{equation}

\noindent where $F'_{0K}$ is the $k$th template spectrum.

The determination of the most probable template spectrum-reddening combination is performed considering two 
degrees of freedom: $j$ and $k$. Given the nature of the problem, it is possible to fix $k$ and vary $j$ 
until $F'_{j}(\lambda)=F'_{0k}(\lambda)$. The combination of the most adequate $j$ and $k$ values can be 
found by minimizing $\chi_{jk}^2$.

Following this procedure it is not difficult at all to minimize the $\chi_{jk}^2$ function  by fixing $k$ and 
varying $j$. We will have a minimum $\chi_{jk}^2$ value within the $E(B-V)_j$ domain searched for each $k$. 
The same procedure can be applied to all available $k$; i.e., for all the existing template spectra, we 
will be able to characterize each $k$ with a $\chi_{jk}^2$ value and finally choose the smallest of the 
$\chi_{jk}^2$ values among them. Although this method appears easy to apply, there remains the problem of selecting the initial value  
for each $k$ from which we start the search for the $\chi_{jk}^2$ minimum. It is therefore 
possible to search the reddening interval from a very negative to a very positive value and to adopt the negative 
value that minimizes the function. However, we have to search with a fine enough screen to make sure that there is no other minimum in the interval. When a large sample of template spectra is handled, the described procedure 
may be rather slow. 

A better approach is to use a combination of both the average and the $\chi^2$  methods. Indeed, the average 
method involves the evaluation of at least one or two members of the sequence of equation (13) of the 
$\chi^2$ method. Thus, an $E(B-V)$ value is quickly obtained, which serves as a starting point to look for the 
$\chi^2$ minimum. It is only necessary to search for the minimum of the $\chi_{jk}^2$ function within a small 
interval ranging from $\left ( <E(B-V)> - \delta \right )$  to $\left (<E(B-V)> + \delta \right )$ with a 
small $\delta$ value.

What will happen if one or more incorrect template spectra have been chosen? In this case, the reddenings resulting 
both from the average and the $\chi^2$  methods will also be wrong, but this is not a problem because, even though 
those templates can be well characterized by a reddening value found by the average method, when the $\chi_{jk}^2$ 
minimum among all templates is found, the mistaken ones will be ignored. We should not expect that a template 
yielding mistaken results when the average method is used could be characterized by a $\chi^2$ value lower than the 
template that would lead to right results. The only relevant templates are those for which the average method works 
correctly. It is precisely in these spectra in which the minimum $\chi_{jk}^2$ function will be found.

The method finally adopted in this work to determine reddening and age of small angular diameter OCs from their 
integrated spectra has the following steps: (i) Normalize the observed spectrum 
and that of the template at $<\lambda>=(\lambda_{max}-\lambda_{min})/2$  so that the number of points on both 
sides of the normalizing point is the same. (ii) Compute the extinction curve 
$\gamma'_{\lambda}=\gamma_{\lambda}-\gamma_{<\lambda>}$. (iii) Determine  
$E(B-V)_i=-ln(G_{\lambda})/\gamma'_{\lambda}$  for each $\lambda$, where $G_{\lambda}$ is the ratio between the normalized 
observed and template spectra. (iv) Determine $<E(B-V)>$ and adopt this value as the starting point for 
evaluating the function $\chi^{2}(E(B-V)_{j})$. (v) Adopt an interval between $<E(B-V)>+\delta$ and $<E(B-V)>-\delta$, 
wherein $\delta$ may be, for example, nearly equal to 1. (vi) Adopt the $E(B-V)_j$ value that 
minimizes function $\chi_{j}^2$  as the cluster reddening. (vii) Finally, follow steps (i) to (vi) for each template spectrum and adopt, as the most 
probable cluster age and reddening, the $j$-th reddening and the age of the $k$-th template spectrum that 
globally minimize the $\chi_{jk}^2$ function.

This mixed procedure is much faster than simply applying the $\chi^2$ method since the search 
interval can be much smaller. It is also more reliable because we can make sure that the $E(B-V)$ that results 
from the search is quite close to the $E(B-V)$ expected from the average method. On the other hand, the mixed 
procedure is more reliable than just the average method, because the search for the minimum of the $\chi_{jk}^2$ 
function involves a global analysis of the spectrum, so localized effects  do not have an influence on the 
final result. Such localized effects can be really troublesome when determining mean reddening. For these 
reasons, we believe that the mixed procedure should be the preferred method to use in characterizing a very large sample 
of template spectra and for determining ages and reddenings of small angular diameter OCs. The computational 
implementation  of the method is relatively simple and allows one to fully systematize and automate age and 
reddening determination of small angular diameter OCs using the existing integrated spectrum template
libraries.

\subsection{Implementation of the Methods}

The full implementation of the method described  was done by developing the computational tool Fast Integrated 
Spectra Analyzer (FISA). We release this application with its complete documentation at: 
http://sites.google.com/site/intspectroscopy. 

Given an OC integrated spectrum covering the 3600 \AA \ $< \lambda <$ 7400 \AA \ spectral range, FISA 
permits a fast and reasonably accurate determination of the cluster age and reddening by using the ''average'' and 
$\chi^2$ methods. The Piatti et al. (2002) and Ahumada et al. (2007) spectral libraries are defined for this 
spectral domain, and these libraries are the ones used by FISA.

If the integrated spectrum in the optical range of an OC is available, FISA makes it relatively easy to quickly 
find the most probable age and reddening values. Basically, FISA works by either manually 
or automatically characterizing the observed spectrum with a reddening that minimizes the $\chi^2$ function for each available 
template spectrum. FISA then selects, from among all the template spectra, the one that globally minimizes $\chi^2$. The 
results are the most probable values of reddening and age for the observed cluster.

Both the average and  the $\chi^2$ methods demand different operations between the observed spectrum and the 
selected template. These operations must be well defined for each wavelength. In general, the wavelength sampling 
of both spectra is not the same, so one must determine the template spectrum flux value 
in the observed spectrum wavelengths. This is done using cubic splines (e.g., Press et al. 2002) which permit 
the interpolation of the nominal flux value quite well. This procedure produces spectra that permit operations among them.

The template integrated spectra  of \citet{pietal02} and \citet{aetal07} 
have been reddening-corrected according to Seaton's (1979) law, so we use this same law in this work. Seaton's law 
includes only a few measured points in the visible region; hence, it is necessary to derive by interpolation a 
smooth extinction curve covering the region where the original spectrum was observed, which is also achieved 
using cubic splines. 

\section{Practical examples}

We now demonstrate the use of FISA through two practical examples. For the first we use observations of the OC 
ESO 502-SC19, which is located at a high Galactic latitude so that very small or nonexistent reddening is 
expected. For the second example, we use observations of the reddened OC Hogg\,22 and then we apply FISA to 
see how well the reddening is recovered. This second example shows how FISA works for a reddened cluster.

\subsection{ESO\,502-SC19}

ESO\,502-SC19 has been studied by \citet{betal01} as one of a number of possibly dissolving OCs. They used 
the reddening maps published by \citet{setal98} to derive $E(B-V) = 0.05$ and from measurements 
made on Digitized Sky Survey (DSS) images, determined such cluster parameters as Galactic coordinates and the 
apparent diameters of both the cluster's major and minor axes. Other than this work, ESO\,502-SC19 has 
not been studied previously, either photometrically or spectroscopically.

The integrated spectrum of ESO\,502-SC19 exhibits the features typical of an intermediate-age cluster. 
Using FISA, the best-matching template is \citet{pietal02} Ia, aged 1 Gyr. ESO\,502-SC19 is found to be unaffected by reddening when the template-matching method is used, in good agreement with the small 
$E(B-V)$ value indicated by Schlegel et al.'s (1998) maps. The reddening distribution used to compute the 
mean reddening required by the average method and the expected theoretical curves are shown 
in Figure 4. The asymmetric systematic deviations at both sides of the normalizing point cause the 
derived mean reddening to deviate from the most probable reddening for ESO\,502-SC19. It is for this reason 
that, starting from the mean reddening, FISA computes the minimum value of the $\chi^2$ function, and 
thus finds the most probable $E(B-V)$ value. This procedure is performed for all available template spectra, 
choosing the one that globally minimizes the $\chi^2$ function, namely the Ia template with $E(B-V) = 
0$. The observed spectrum, the adopted 
template and the residual flux between both are shown in Figure 5.

\subsection{Hogg\,22}

The sparse cluster Hogg\,22 has been studied photometrically and spectroscopically. A compilation and discussion 
of these results can be seen in Ahumada et al. (2007). They showed that there is good agreement between the 
integrated spectrum of Hogg\,22 - neglecting the contribution of the bright star HD\,150958 and applying a reddening 
of $E(B-V) = 0.55$ - and the template spectrum Yb1. Using FISA, one finds the integrated spectrum is again well fit 
with template Yb1 and with a similar reddening of $E(B-V) = 0.58$. The reddening results from the average 
method are shown in Figure 6. In spite of some systematic deviations, the mean reddening 
 that resulted [$<E(B-V)> \sim 0.52$] is very close to the value finally adopted. Figure 7 shows the template 
spectrum Yb1, the observed spectrum of Hogg\,22, that spectrum corrected for the adopted 
reddening [$E(B-V) \sim 0.58$] and residuals in flux between the template and the corrected spectra.

\section{Conclusions}

Over more than a decade, the integrated spectroscopic technique has been developed and successfully applied 
at the Observatorio Astron\'omico of the Universidad Nacional de C\'ordoba (Argentina). Its application to 
different astronomical objects has led to interesting results in relation to OCs (e.g., Ahumada et 
al. 2001), globular clusters (e.g., Bica et al. 1998), planetary nebula (e.g., Bica et al. 1995), galaxies 
(e.g., Dutra et al. 2001) and even supernova remnants (e.g., Bica et al. 1995). However, during recent years, 
the technique has been mostly used to determine the fundamental properties of compact star clusters belonging 
both to our Galaxy (e.g., Ahumada et al. 2000) and to the Magellanic Clouds (e.g., Santos et al. 2006).
 
The main aim of this article was to describe a  method in detail that not only allows us to systematize and 
automate the analysis of a large amount of integrated spectra, but also to obtain astrophysical results. 
We started by demonstrating in \S 2 that the integrated radiation distribution of a stellar population 
is determined by its stage of evolution. Since OCs can be considered true stellar population units, because 
all their stars formed at the same time and they have about the same metallicity, their integrated spectra 
provide us with information about their ages. We then described three different approaches to the characterization 
of a cluster age from its integrated spectrum. We also presented some arguments validating the use of template 
spectrum libraries for interpreting an observed spectrum. As the existing dispersion in Galactic OC metallicities 
is small - especially in young clusters - these stellar systems stand out as excellent candidates to be observed 
through integrated spectroscopy and to be analyzed using template spectra. 
 
We presented in \S 3 two different methods based on integrated spectroscopy to estimate age and reddening 
of small angular diameter OCs. The average method is quite useful as a starting point to estimate the 
reddening affecting a cluster. It was shown, however, that this method is not reliable by itself, since many systematic differences between the observed and the template spectra are to be 
expected when put into practice. We demonstrated that the errors introduced by these  systematic effects can be removed by using the 
$\chi^2$  method, which takes into account the overall spectrum shape so that variations over a small spectral 
range do not have a marked influence. We concluded that the safest option for estimating a cluster reddening 
from integrated spectra is to use a combination of both the average and $\chi^2$ methods. The use of 
the average method permits a significant acceleration of the required calculations so that this approach is 
appropriate when dealing with a great number of clusters. Finally, we showed how this method can be practically 
implemented by deriving an extinction curve for the spectral range of the observed spectrum and interpolating 
the template spectrum data to obtain values at the measured points in the observed spectrum.
 
The theory developed in \S 2, as well as the algorithms presented in \S 3, allowed us to build a 
computational application that greatly facilitates the age and reddening determination from integrated 
spectra of small angular diameter star clusters through the use of currently available integrated spectral 
libraries. We illustrated the use of this new tool, which we called ''Fast Integrated Spectra Analyzer'' (FISA), 
with two examples. We believe FISA represents a step forward in the formerly time-consuming task of analyzing 
integrated spectra. From now on, FISA will enable us to analyze large amounts of integrated 
spectroscopic data in relatively short times, thus simplifying the determination of age and reddening of 
small angular diameter OCs. In a forthcoming article, we will present the results obtained from integrated spectra 
obtained at CASLEO for about 50 OCs whose features are presently poorly known or practically 
unknown.

\begin{deluxetable}{l l l l l l }
\tablewidth{0pt}
\tablecaption{LIBRARY OF TEMPLATE SPECTRA TAKEN FROM PIATTI ET AL. (2002) AND AHUMADA ET AL. (2007).}
\tablehead{
                         
\colhead{Name} & \colhead{Age range} & \colhead{Members} & \colhead{Name} & \colhead{Age range} 
& \colhead{Members} \\
 \colhead{} & \colhead{(Myr)} & \colhead{} & \colhead{} & \colhead{(Myr)} & \colhead{}
}
\startdata

   Ya1  & 2-4 & 3 &   Yb2\_WR  & & 6 \\
   Ya2  & &  &  Yb3\_WR & & 6 \\
   Ya3 & & &Yc  &20& 2 \\
   Yab & 4-5 & 2 &  Ycd & 30 & 3 \\   
   Ya1\_WR  & & 5 &Yd  &40 & 3 \\ 
   Ya2\_WR  & &  &   Ye &45-75 & 5 \\
   Ya3\_WR & & &     Yf  &100-150 & 6 \\
   Yb1  & 5-10 & 3 &   Yg  &200-350 & 7 \\
   Yb2  & &  &     Yh &500& 2 \\
   Yb3 & & &    Ia  &1000 & 4 \\
   Yb1\_WR  & & 6 & Ib  &3000-4000 & 3 \\

\enddata

\tablecomments{ These 
templates are loaded in FISA. All template spectra have solar metallicity. Yc and Ib templates were redefined in 
Ahumada et al. (2007).}

\
\

\end{deluxetable}

\begin{figure}
\centering
\includegraphics[width=15cm]{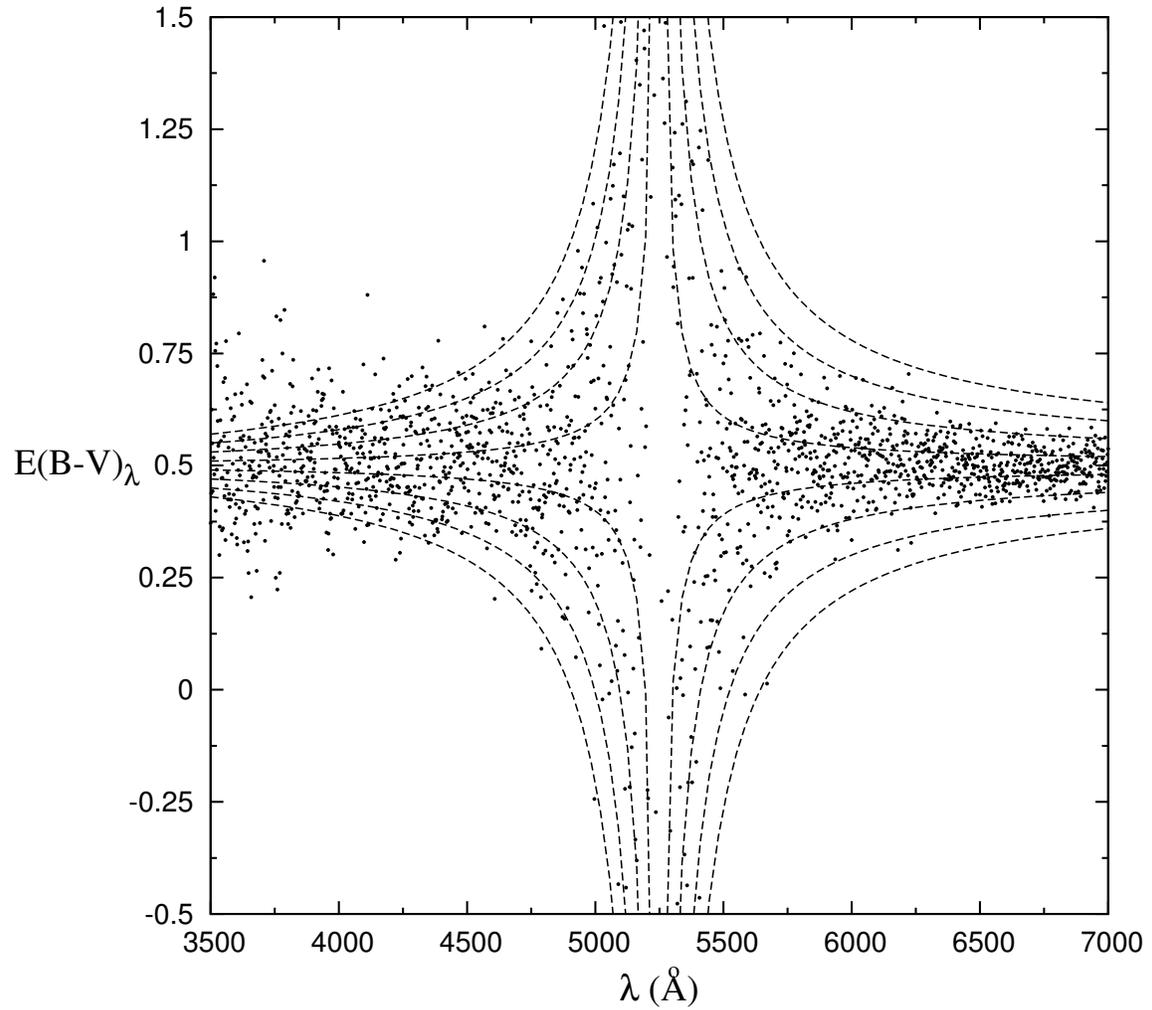}
   \caption{Derived $E(B-V)$ values (points) and four of the many possible theoretical distributions 
(dashed lines). The normalization wavelength is 5248 \AA. The large scatter on the left is due to deviations 
from the standard $1/\lambda$ extinction curve.}
\end{figure}

\begin{figure}
\centering
\includegraphics[width=15cm]{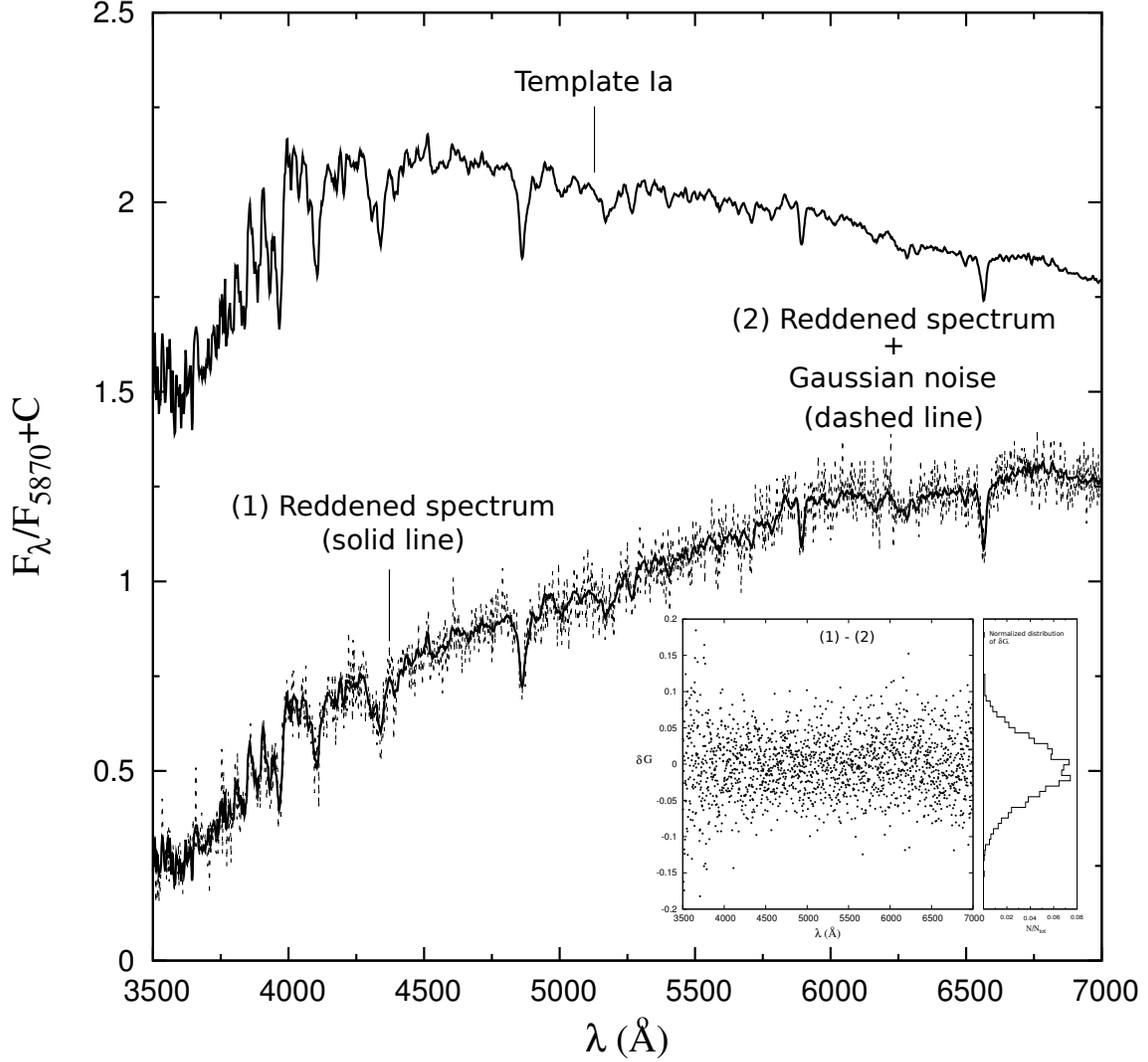}
 \caption{From top to bottom: The Ia template spectrum and the same spectrum now reddened by $E(B-V) = 0.5$ 
(solid line).  Overlaid on the reddened spectrum is the spectrum that results from adding noise with a wavelength-independent standard deviation (dashed line). Also shown is the distribution of $\delta G$, defined as the difference 
between the fluxes of the reddened spectrum without noise and the one with noise added ({\it small box}).}
\end{figure}

\begin{figure}
\centering
\includegraphics[width=15cm]{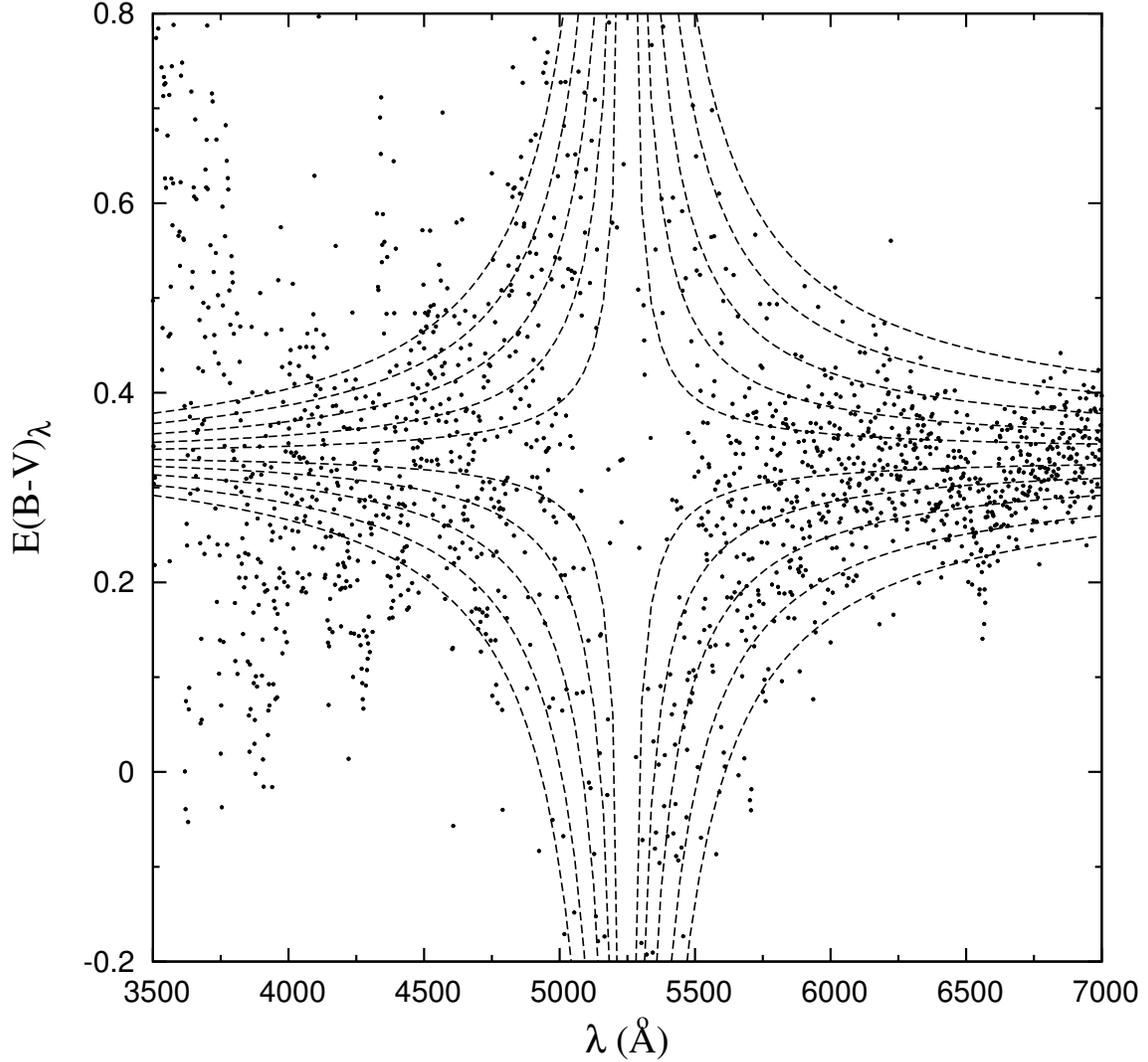}
   \caption{ Distribution of computed $E(B-V)$ values (points) when a poorly matching template spectrum is used. 
   The large and systematic deviations with respect to the theoretically expected curves (dashed lines) are mainly the 
result of the observed and template spectra having different spectral features.  Thus, the differences between the 
 observed and the template spectra are not being determined  by the extinction.}

\end{figure}

\begin{figure}
\centering
\includegraphics[width=15cm]{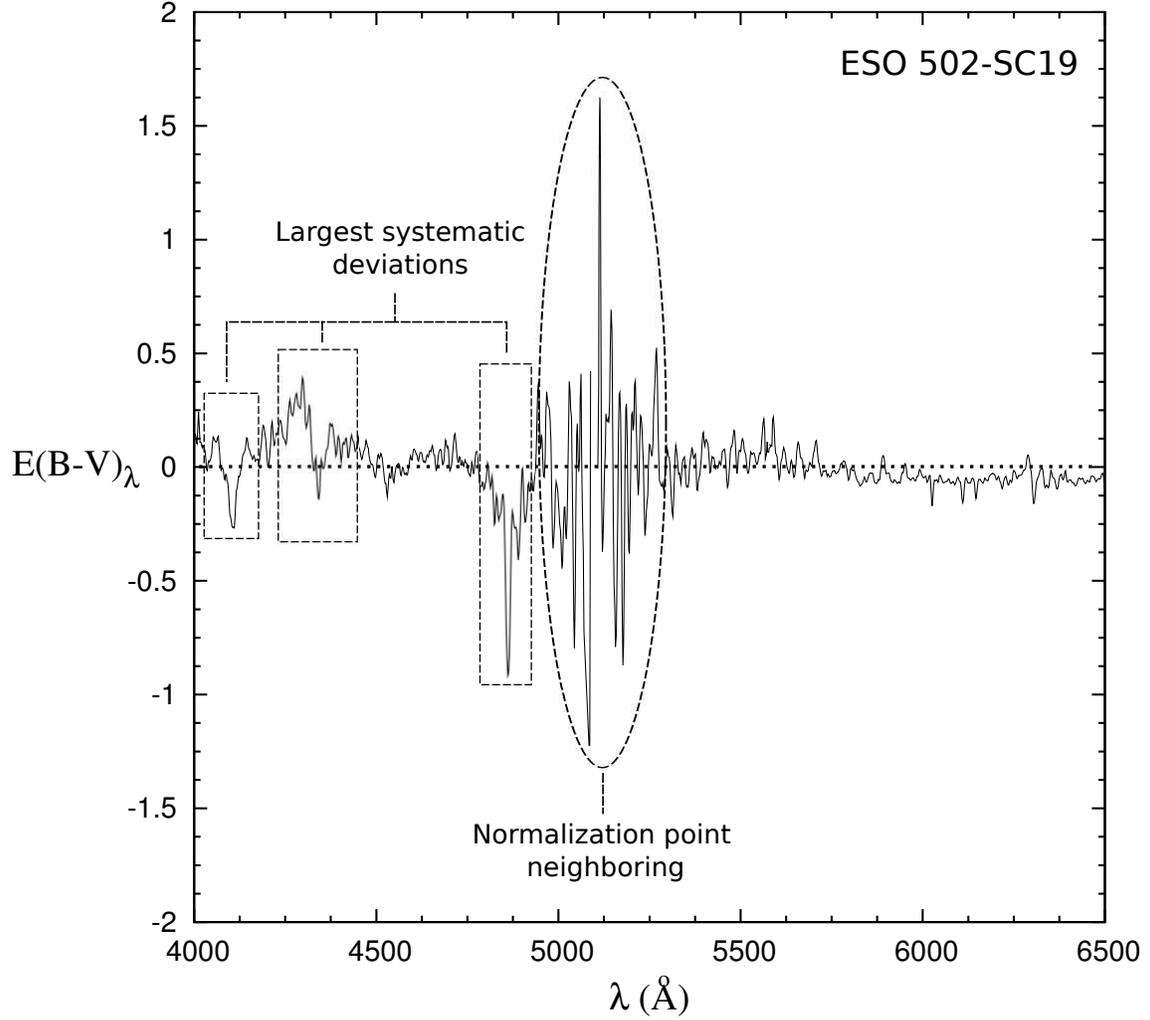}
   \caption{Distribution of $E(B-V)$ values from comparing the observed spectrum with the Ia template 
of Piatti et al. (2002).  The systematic deviations at both sides of the normalizing point, due largely to spectral 
features that differ in the two spectra, cause the derived mean reddening to differ slightly from the most 
probable reddening value adopted for ESO\,502-SC19. }

\end{figure}

\begin{figure}
\centering
\includegraphics[width=15cm]{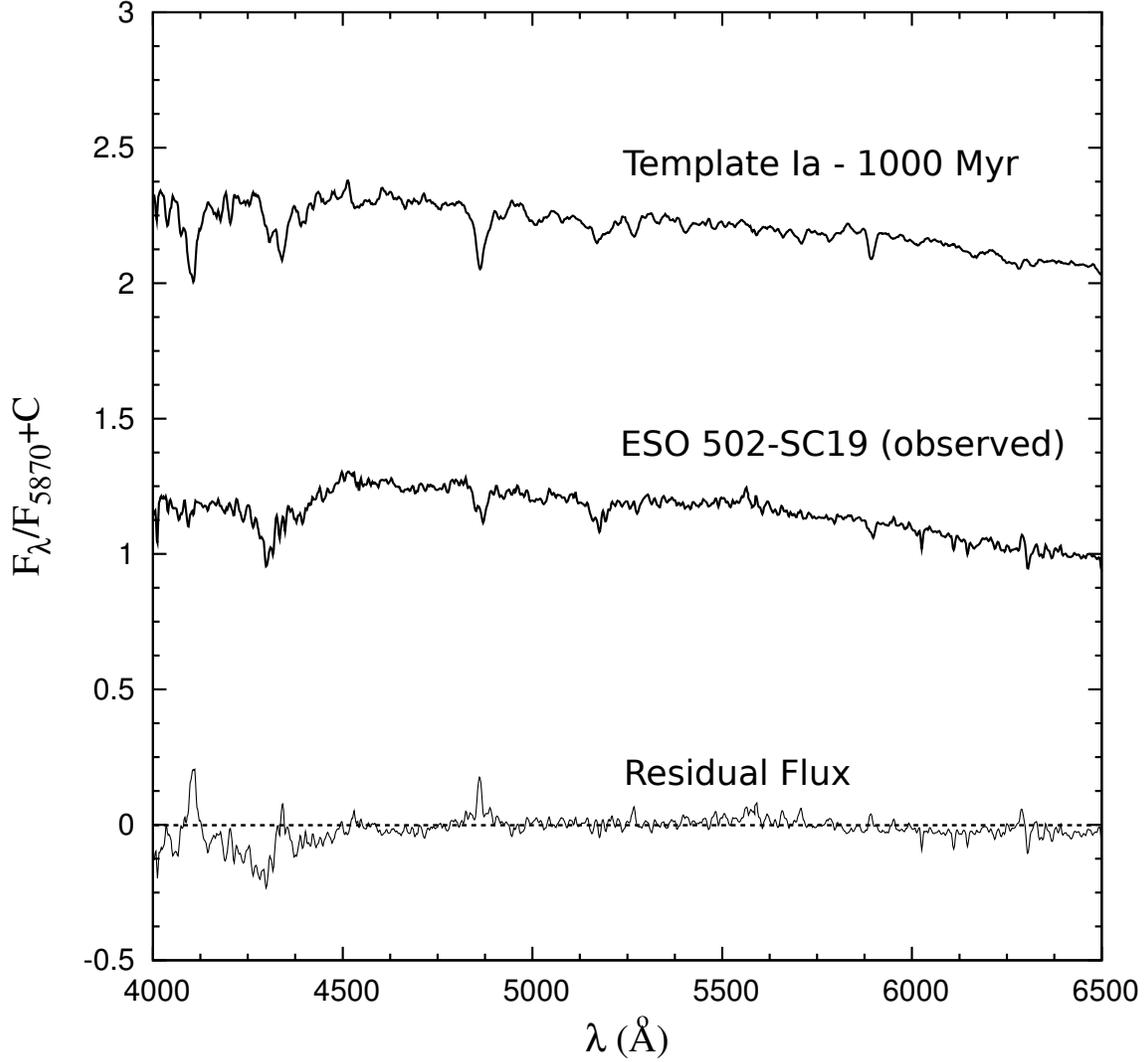}
   \caption{From top to bottom: The selected template spectrum, the spectrum of ESO\,502-SC19 and the difference in 
   flux difference  between them computed through $(F_{cluster}-F_{template})/F_{cluster}$. Note the similarity between 
   the larger values of flux differences and the anomalous $E(B-V)$ values in Fig. 4.}
\end{figure}

\begin{figure}
\centering
\includegraphics[width=15cm]{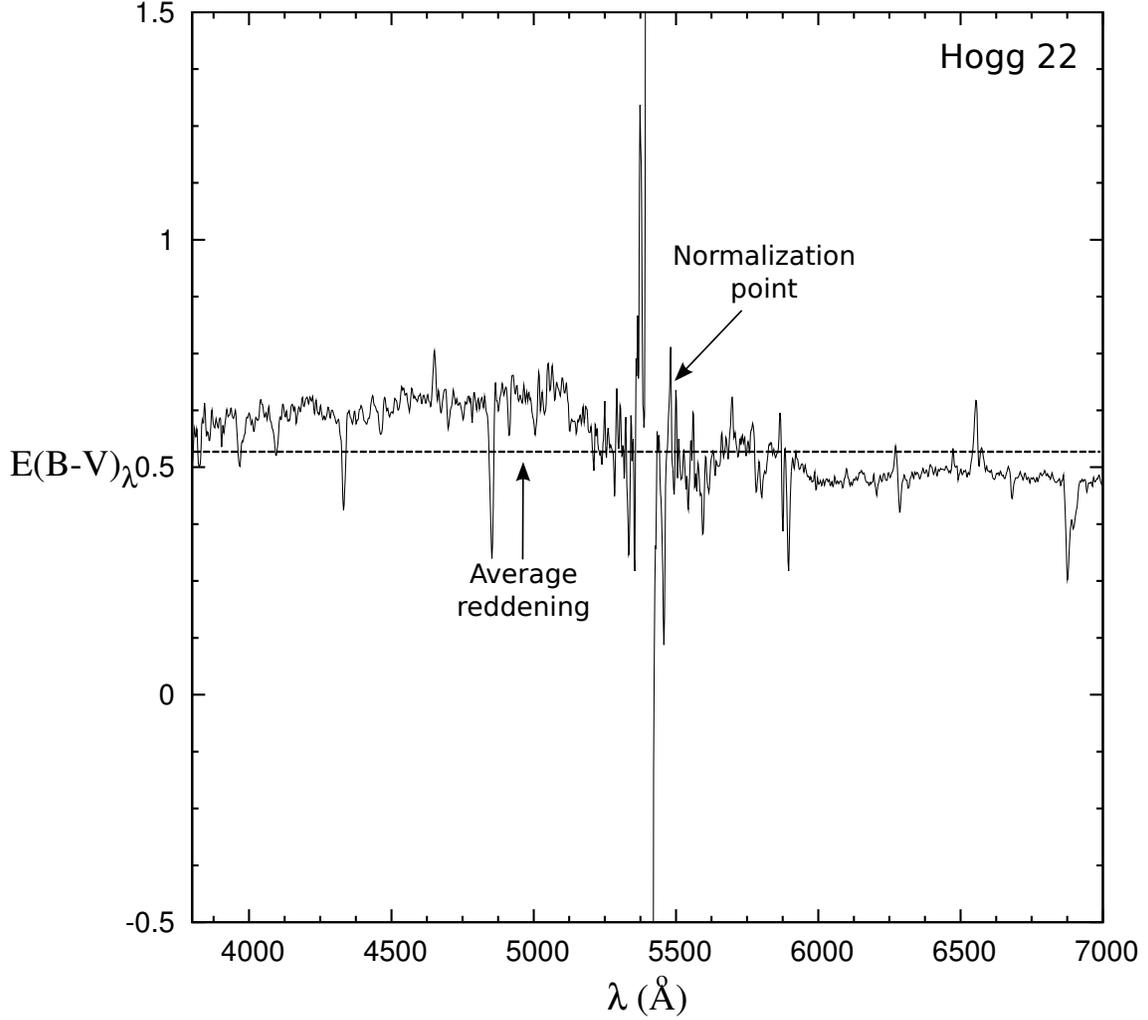}
   \caption{Distribution of the computed $E(B-V)$ values from comparing the observed spectrum of Hogg\,22 with the Yb1 template 
spectrum of Piatti et al. (2002). As expected, there are large deviations near the normalization point. Despite this and the  
variations over the wavelength range, the reddening seems to be very well determined from the mean value 
produced by the average method.}
\end{figure}

\begin{figure}
\centering
\includegraphics[width=15cm]{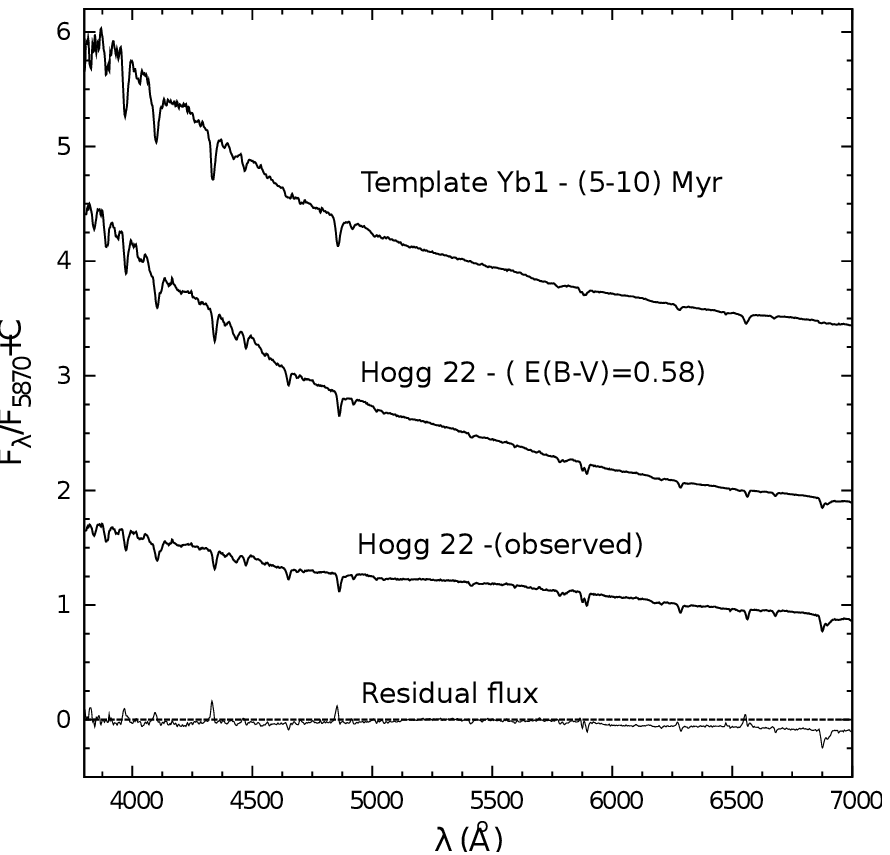}
   \caption{From top to bottom: The chosen template spectrum Yb1, the Hogg\,22 spectrum dereddened by 
$E(B-V)$ = 0.58, the original (observed) Hogg\,22 spectrum  and the flux difference between the dereddened 
spectrum and the template spectrum computed through $(F_{cluster}-F_{template})/F_{cluster}$.}
\end{figure}

\acknowledgments

We thank the staff and personnel at CASLEO for hospitality and assistance during the observations. We would like to 
thank the anonymous referee for his or her helpful comments and suggestions that contributed to the improvement of the article. We also are indebted to W. Osborn, Yerkes Observatory, for carefully reading this manuscript and 
making some suggestions. The authors acknowledge use 
of the $CCD$ and data acquisition system supported under NSF grant AST-90-15827 to R. M. Rich. We 
gratefully acknowledge financial support from the Argentinian institutions CONICET, FONCyT and SECyT (Universidad Nacional de 
C\'ordoba).\\

{}

\end{document}